\documentclass[10pt,letterpaper,twocolumn]{article}

\usepackage{ol2}
\usepackage{amsmath}

\begin{document}

\twocolumn[

\title{Chaotic dynamics of frequency combs generated with continuously pumped nonlinear microresonators}

\author{A. B. Matsko$^*$, W. Liang, A. A. Savchenkov, and L. Maleki}

\address{OEwaves Inc., 465 North Halstead Street, Suite 140, Pasadena, CA 91107 \\
$^*$Corresponding author: andrey.matsko@oewaves.com }

\begin{abstract}
We  theoretically and experimentally investigate the chaotic regime of optical frequency combs generated in  nonlinear ring microresonators pumped with continuous wave light. We show that the chaotic regime reveals itself, in an apparently counter-intuitive way, by a flat top symmetric envelope of the frequency spectrum, when observed by means of an optical spectrum analyzer. The comb demodulated on a fast photodiode  produces a noisy radio frequency signal  with an spectral width significantly exceeding the linear bandwidth of the microresonator mode.
\end{abstract}

\ocis{190.3100, 190.2620, 190.4223, 140.4780, 350.3950}

 ]

Oscillators are fundamentally nonlinear systems with intrinsically rich dynamics.  They exhibit chaotic behavior, or stable oscillation, based on the choice of their operating parameters.  In the case of optical oscillators, any and all examples of dynamical behavior can be, and have been, observed.  In particular, onset of oscillation in mode locked laser has been known to be associated with both hard and soft excitation, determined by the dynamics of the system.  Soft excitation refers to oscillation that self starts due to the presence of noise in the oscillator circuit.  Hard excitation refers to a regime whereby an external impulse is required to start the oscillation.  In this Letter we investigate both chaotic and stable regimes of oscillation in frequency combs generated in monolithic ring microresonators made with transparent materials possessing Kerr nonlinearity \cite{delhaye07n}.  These optical combs have been the subject of intense interest lately in connection with applications in a variety of areas in science and technology \cite{kippenberg11s}.  However, a satisfactory answer to the question of coherence of these combs, especially for those exhibiting a wide frequency spectrum, has been elusive. We show that a large number of optical Kerr combs reported in the literature are in fact chaotic, despite their evidently regular and symmetric shape.  These chaotic combs will naturally display a noisy RF spectrum when demodulated on a fast photodiode, as has been reported in various studies.

The existence of chaotic regime is consistent with the intrinsic nature of Kerr frequency combs.  What is surprising is that a comb exhibiting a regular spectral envelope with a flat top, and with modes symmetrically positioned about the pump, can be chaotic.  As shown below, the observed shape of the comb is in fact the result of time-averaged chaos, and is the consequence of the conventional means for observing these combs, namely the use of an optical spectrum analyzer (OSA).
\begin{figure}[ht]
  \centering
  \includegraphics[width=8.5cm]{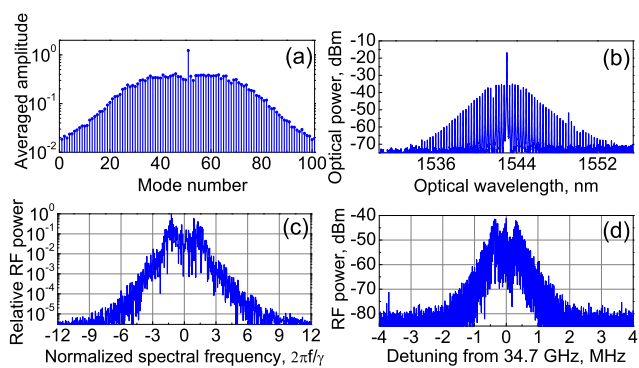}
\caption{ Optical spectrum of an incoherent Kerr comb ((a) and (b)) and associated RF signals ((c) and (d)) generated by the comb on a fast photodiode. Parts (a) and (c) represent the theoretical simulation results, (b) and (d) are for the measurement results. To simulate the spectrum of the chaotic comb (a) we time averaged the instantaneous chaotic comb spectra over $t_{ave} = 100/\gamma$. This averaging corresponds to the averaging introduced by an OSA used to observe the comb spectrum in the experiment.
} \label{fig1}
\end{figure}

The observed incoherent frequency comb generation in a magnesium fluoride (MgF$_2$) whispering gallery mode (WGM) microresonator, and results of numerical simulation reproducing the experimental data are shown in Fig.~\ref{fig1}. The envelope has a characteristic flat top distinct from envelopes of other combs that also can be generated in the system. We found that the RF signal produced on a fast photodiode by this comb has a spectral width that significantly exceeding the linewidth of the resonator modes, resulting from the chaotic dynamics of light confined in the microresonator. The presence of chaos is more general than the nonlinear interaction of  subcombs \cite{herr12np} since it constitutes a fundamental property of the nonlinear oscillator described with forced nonlinear Schrodinger equation \cite{blow84prl}.

The existence of a chaotic comb does not imply that the resonator cannot generate a coherent Kerr comb. A stable oscillation mode can readily exist in a comb oscillator and may be observed. This, however, is not trivial since the majority of  regular oscillation solutions are associated with hard excitation regime \cite{matsko12pra}. We show that an abrupt switching of frequency of the cw pump laser facilitates transforming the chaotic comb regime to a regular one.
\begin{figure}[ht]
  \centering
  \includegraphics[width=7.cm]{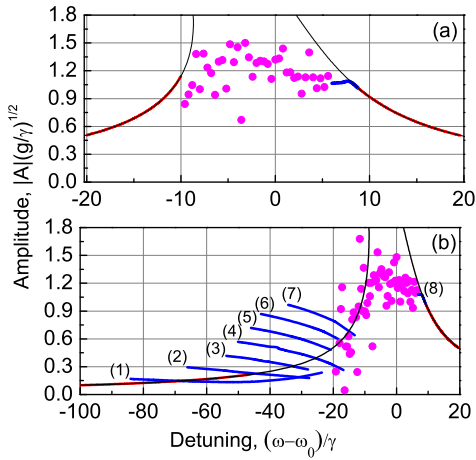}
\caption{ Dependence of the normalized field amplitude ($A=b_{51}$, for the case described in the paper) in the optically pumped mode on the frequency of the cw pump light. Solid lines show stable branches corresponding to generation of various Kerr combs (see Fig.~\ref{fig3}), dots correspond to generation of chaotic combs or combs with time-depending envelopes. Plot (a) stands for the soft excitation regime \cite{matsko12pra}. The initial amplitude in this case is selected to be $|b_j|(g/\gamma)^{1/2}=10^{-4}$, which approximately corresponds to the field in the mode containing $1/2$ photon. Branches (1-7) are not accessible in this case. Plot (b) stands for hard excitation. It is assumed that $A(g/\gamma)^{1/2}=10$ and $b_{50}(g/\gamma)^{1/2}=0.3 \dots 3$ and $b_{52}(g/\gamma)^{1/2}=0.3 \dots 3$ to get the solutions. Stable solutions are realized for the particular nonzero initial condition and detuning selection. The comb can also be excited from nearly zero initial excitation of modes, if the pump frequency is shifted  nonadiabatically, as shown in Fig.~\ref{fig4}. Constant tuning of the pump frequency also allows observing solutions corresponding to stable branches \cite{herr12arxiv,note}.
} \label{fig2}
\end{figure}

We generated a chaotic Kerr comb in a MgF$_2$ WGM microresonator. The resonator had 34.7~GHz free spectral range (resonator radius was approximately equal to $1$~mm), and a loaded Q-factor of about $10^9$.  The resonator was produced from a commercially available z-cut MgF$_2$ optical window by mechanical polishing. The shape of the resonator was that of an oblate spheroid in the vicinity of WGM localization.  Two evanescent field prism couplers were used with the resonator\cite{maleki09chapter}. The pump light was injected through one prism and the Kerr comb was sampled using the other prism. In this way we observed the comb structure excited within the resonator and avoided excessive leakage of the pump light to the photodiode used to generate RF signal. The resonator was pumped with a 1543~nm distributed feedback diode laser self-injection locked \cite{liang10ol} to an appropriate WGM mode. The locking occurred due to the intrinsic Rayleigh scattering in the resonators host material. The optical power emitted by the laser was approximately 5~mW. The input coupling efficiency was approximately 30\%, and approximately 50\% of the light exiting the second prism coupler was collected into a multimode fiber and sent either to an optical spectrum analyzer, or a fast photodiode and an RF spectrum analyzer.

An example of the observed optical and RF spectra is shown in Figs.~\ref{fig1}(b) and \ref{fig1}(d). While the optical spectrum is rather regular, the RF spectrum is very wide. It is approximately five times wider than the mode bandwidth (the bandwidth of the WGMs is approximately 200~kHz). This implies that the frequency comb is incoherent.
\begin{figure}[ht]
  \centering
  \includegraphics[width=8.5cm]{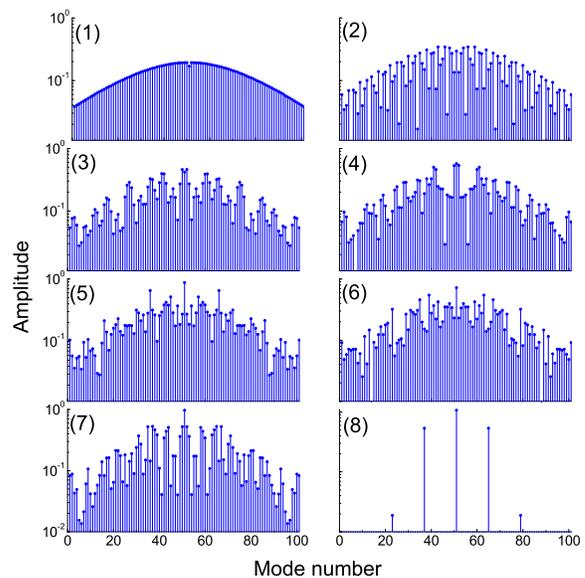}
\caption{ Simulated spectra of the combs corresponding to stable solution branches shown in Fig.~\ref{fig2}.
} \label{fig3}
\end{figure}

We performed numerical simulations along the lines discussed in \cite{matsko12pra,chembo10pra}. We introduced an interaction Hamiltonian $\hat V= -\hbar g (\hat e^\dag)^2 \hat e^2/2$, where $g=\hbar \omega_0^2 c n_2/({\cal V}n_0^2)$ is the coupling parameter obtained under the assumption of complete space overlap of the resonator modes \cite{matsko05pra}, $\omega_0$ is the value of the optical frequency of the externally pumped mode, $c$ is the speed of light in vacuum, $n_2$ is the cubic nonlinearity of the material, ${\cal V}$ is the effective geometrical volume occupied by the optical modes in the resonator, and $n_0$ is the linear index of refraction of the resonator host material. The operator $\hat e$ is given by the sum of annihilation operators of the electromagnetic field for all the interacting resonator modes that we took into consideration $\hat e= \sum \hat b_j$.

Equations describing the evolution of the field in the resonator modes are generated using the
Hamiltonian
\begin{equation} \label{set}
\dot{\hat b}_j=-(\gamma+i\omega_j)\hat b_j+ \frac{i}{\hbar} [\hat V,\hat b_j]+F e^{-i\omega t}
\delta_{j0,j},
\end{equation}
where $\delta_{j0,j}$ is the Kronecker's delta; $j_0$ is the number of the externally pumped mode, $\gamma$ is the half width at the half maximum (HWHM) for the optical modes, assumed to be the same for the all modes involved. The external pump is applied to the central mode of the group of 101 modes. The pump amplitude is given by $F= [(2 \gamma_{0c} P)/(\hbar \omega_0)]^{1/2}$, where $P$ is the value of the power of the pump light. We studied the case of pump power given by $F(g/\gamma)^{1/2}=10$, and anomalous GVD set by $2\omega_0-  \omega_{+}-  \omega_{-} = -0.1\gamma$.

The results of  simulations is shown in Figs.~\ref{fig1}(a) and \ref{fig1}(c). To obtain the optical spectrum we employed the fact that a typical OSA time averages each frequency component. Therefore, the instantaneous frequency comb envelope in not the observable in the experiment, rather it is the comb averaged for times longer than 10~ms that is observed. The instantaneous envelope of the comb looks chaotic and not symmetric, but the averaged comb looks regular and similar to the experimental observation. The simulated RF signal is also similar to the measured RF signal, and is broader compared with the HWHM of the modes.

\begin{figure}[ht]
  \centering
  \includegraphics[width=8.cm]{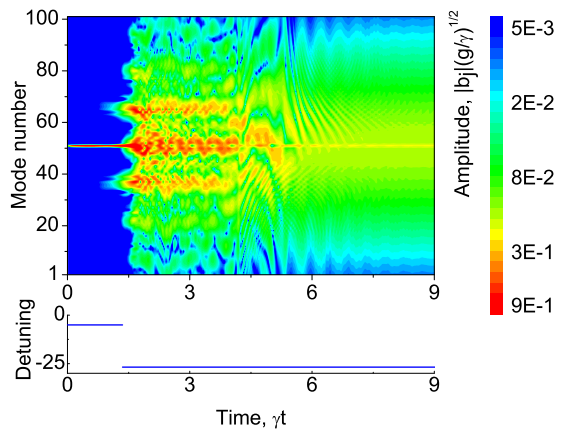}
\caption{ A density plot illustrating time dependence of amplitude of comb harmonics. This is an example of excitation of a regular frequency comb in an empty microresonator via nonadiabatic change of the oscillator parameters. The comb forms starting at $|b_j(t=0)|=10^{-4}$ when the normalized frequency detuning ($(\omega-\omega_0)/\gamma$) between the pump light and the pumped mode abruptly changes from $-5$ to $-27$ at $t=1.4/\gamma$. A chaotic frequency comb is generated if the detuning stays constant at the initial level ($-5$). The abrupt change of the detuning results in generation of a regular broad coherent comb at branch (1) of Fig.~\ref{fig2}. The transient process reveals breather formation in the resonator.
} \label{fig4}
\end{figure}

We studied regimes of  comb generation numerically and found that the chaotic regime is always accessible, independent of the initial conditions, Fig.~\ref{fig2} (soft excitation regime \cite{matsko12pra}). Generation of regular combs (Fig.~\ref{fig3}), on the other hand, depends on the initial conditions and cannot be realized if the resonator is initially empty (hard excitation regime \cite{matsko12pra}). To achieve steady comb generation in an initially empty resonator one needs to nonadiabatically change either the power or the detuning of the pump light, the two parameters that are accessible in the comb oscillator. We demonstrated that an abrupt change of the pump frequency result in formation of a stable frequency comb Fig.~\ref{fig4}. The comb is started from chaotic oscillations, but when the laser frequency moves to to the position where regular solutions are allowed, the oscillation pattern changes. The modification of the oscillation regime is accompanied by intricate comb dynamics. This is not the only way for generation of an stable comb. It was shown in \cite{herr12arxiv,note} that scanning the pump frequency through the resonator mode also allows achieving stable combs.

To conclude, we have demonstrated, both experimentally and theoretically, the generation of chaotic Kerr combs in a nonlinear microresonator. We provide explanation for why in many experiments the comb spectrum is often observed to be symmetric, implying the presence of high coherence, while the RF beatnote produced by the comb is completely irregular, implying an incoherent comb. We have shown that it is possible to tune the comb oscillator from the chaotic regime to the regular oscillation by a nonadiabatic change of frequency or power of the pump laser.

The authors acknowledge support from Defense Sciences Office of Defense Advanced Research Projects Agency under contract No.~W911QX-12-C-0067.

\newpage

\section*{References}

\begin{enumerate}

\item P. Del'Haye, A. Schliesser, O. Arcizet, T. Wilken, R. Holzwarth, and T. J. Kippenberg, "Optical frequency comb generation from a monolithic microresonator," Nature  {\bf 450}, 1214--1217 (2007).

\item T. J. Kippenberg, R. Holzwarth, and S. A. Diddams, "Microresonator-based optical frequency combs," Science {\bf 332}, 555--559 (2011).

\item T. Herr, K. Hartinger, J. Riemensberger, C. Y. Wang, E. Gavartin, R. Holzwarth, M. L. Gorodetsky, and T. J. Kippenberg, "Universal formation dynamics and noise of Kerr-frequency combs in microresonators," Nature Photonics {\bf 6}, 480–-487 (2012).

\item K. J. Blow and N. J. Doran, "Global and Local Chaos in the Pumped Nonlinear Schrodinger Equation," Phys. Rev. Lett. {\bf 52}, 526-529 (1984).

\item A. B. Matsko, A. A. Savchenkov, V. S. Ilchenko, D. Seidel, and L. Maleki, "Hard and soft excitation regimes of Kerr frequency combs," Phys. Rev. A {\bf 85}, 023830 (2012).

\item T. Herr, V. Brasch, M. L. Gorodetsky, T. J. Kippenberg, "Soliton mode-locking in optical microresonators," $arXiv:1211.0733$ (2012).

\item Our paper was at the final preparation stage when \cite{herr12arxiv} was published.

\item L. Maleki, V. S. Ilchenko, A. A. Savchenkov, and A. B. Matsko, "Crystalline Whispering Gallery Mode Resonators in Optics and Photonics," Chapter {\bf 3} in {\em Practical Applications of Microresonators in Optics and Photonics} edited by A. B. Matsko, (CRC Press, 2009).

\item W. Liang, V. S. Ilchenko, A. A. Savchenkov, A. B. Matsko, D. Seidel, and L. Maleki, "Ultra-narrow linewidth external cavity semiconductor lasers using crystalline whispering gallery mode resonators," Opt. Lett. {\bf 35}, 2822--2824 (2010).

\item Y. K. Chembo and N. Yu, "Modal expansion approach to optical-frequency-comb generation with monolithic whispering-gallery-mode resonators," Phys. Rev. A {\bf 82}, 033801 (2010).

\item A. B. Matsko, A. A. Savchenkov, D. Strekalov, V. S. Ilchenko, and L. Maleki, "Optical hyper-parametric oscillations in a whispering gallery mode resonator: threshold and phase diffusion," Phys. Rev. A {\bf 71}, 033804 (2005).

\end{enumerate}


\begin{thebibliography}{99}
\bibitem{note} Our paper was at the final preparation stage when \cite{herr12arxiv} was published.
\end{thebibliography}
\end{document}